\documentclass[aps,prb,showpacs,twocolumn,superscriptaddress]{revtex4-2}

\usepackage{graphicx} 
\usepackage{color}
\usepackage{array}
\usepackage{mathrsfs}

\usepackage{ulem} % sout

\begin{document}
\title{Magnetic phase diagrams of the spin-$\frac{1}{2}$ Heisenberg model on a kagome-strip chain: Emergence of a Haldane phase}

\author{Katsuhiro Morita}
\email[e-mail:]{katsuhiro.morita@rs.tus.ac.jp}
\affiliation{Department of Applied Physics, Tokyo University of Science, Tokyo 125-8585, Japan}
\affiliation{Department of Physics, Faculty of Science and Technology, Tokyo University of Science, Chiba 278-8510, Japan}

\author{Shigetoshi Sota}
\affiliation{RIKEN Center for Computational Science (R-CCS), Kobe, Hyogo 650-0047, Japan}
\affiliation{RIKEN Center for Quantum Computing (RQC), Wako, Saitama 351-0198, Japan}

\author{Takami Tohyama}
\affiliation{Department of Applied Physics, Tokyo University of Science, Tokyo 125-8585, Japan}

\date{\today}

\begin{abstract}
Frustrated one-dimensional quantum spin systems are known to exhibit a variety of quantum ground states due to the effects of quantum fluctuations and frustrations. 
In a spin-1/2 kagome-strip chain, which is one of the frustrated one-dimensional spin systems, many quantum phases have been found. However, the magnetic phase diagrams of the kagome-strip chain under magnetic field have not been fully understood.  We construct magnetic phase diagrams at 0, 1/5, 3/10, 1/3, 2/5, 7/15, 3/5, and 4/5 magnetization ratio in the kagome-strip chain and investigate magnetic properties  in each phase using the density matrix renormalization group method. We find fifteen magnetization-plateau phases, one of which is equivalent to the spin-1 Haldane phase.
\end{abstract}

\maketitle
\section{Introduction}

Quantum spin chains have been studied extensively because they have remarkable quantum features.
The ground state of single Heisenberg chains with the half integer spin is the Tomonaga-Luttinger liquid~\cite{TL} showing gapless excitation, 
while the ground state for the integer spin is the Haldane state with a finite energy gap called the Haldane gap~\cite{Hald}.
In addition, the Heisenberg models describing frustrated quantum spin chains are known to exhibit various types of quantum phases, which have been studied intensively.  
Magnetization plateaus and strong quantum entanglement states due to the competition between frustration and quantum fluctuations have been found in the diamond chain~\cite{DC1,DC2,DC3}, kagome-strip chains (KSCs)~\cite{ksc1,ksc2,ksc3,ksc4,ksc5,ksc6,ksc7,ksc8,ksc9}, and various frustrated one-dimensional chains~\cite{ZZ1,ZZ2,ZZ3,ZZ4,ZZ5, STC1,STC2, TT1,TT2,TT3,TT4,TT5,TT6, FT1,FT2, OT1,OT2,OT3,OT4}.

In two-dimensional frustrated quantum spin systems, the spin-1/2 kagome lattice Heisenberg model,  
which is one of the most famous models, 
exhibits multiple magnetization plateaus at $M/M_{\rm sat}$ = 0, 1/9, 1/3, 5/9, and 7/9, 
where $M$ is the magnetization and $M_{\rm sat}$ is the saturation magnetization~\cite{KMH1,KMH2,KMH3,KMH4,KMH5,KMH6,KMH7}. 
However, our understanding of magnetic properties of the kagome-lattice model is far from complete, because of the difficulty of calculating the frustrated two-dimensional systems.

\begin{figure}[tb]
\includegraphics[width=86mm]{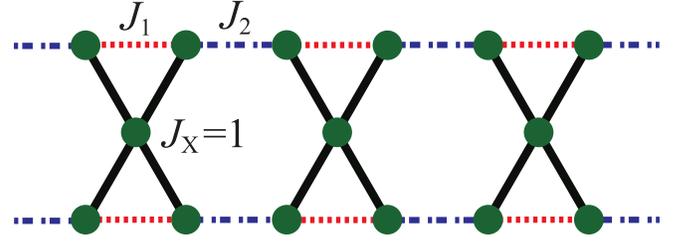}
\caption{Structure of a kagome-strip chain. The black solid, red dashed, and blue broken lines denote the exchange interactions $J_{\rm X}$, $J_1$, and $J_2$, respectively.
The green circles refer to the sites with a spin.
We set $J_{\rm X}$=1 as the energy unit.
\label{lattice}}
\end{figure}

The spin-1/2 KSC shown in Fig.~\ref{lattice}, which has the lattice structure similar to the kagome lattice, exhibits magnetization plateaus at $M/M_{\rm sat}$ = 0, 1/5, 3/10, 1/3, 2/5, 7/15, 3/5, and 4/5~\cite{ksc6}. 
One of the 1/5 plateau phases has features of both the half-integer spin and integer spin chains~\cite{ksc9}.
This implies that the KSC has  strong frustration effects similar to the kagome lattice and
strong quantum fluctuations inherent in one dimension unlike the two-dimensional kagome lattice.
Moreover, to the best of our knowledge, there is no such a spin model that exhibits so many magnetization plateaus and has a phase with features of both the half-integer spin and integer spin chains.
However, except for  $M/M_{\rm sat} = 1/5$, the magnetic phase diagrams of the KSC have not been completely determined. 
Therefore, there is a possibility of finding new phases that have not been found in
previous studies~\cite{ksc6,ksc9}. Constructing complete magnetic phase diagram under the magnetic field will provide a strong motivation to experimentally investigate a model compound of the KSC reported before~\cite{kscexp}.
Therefore, further study of the KSC is required. 

\begin{figure*}[tb]
\includegraphics[width=180mm]{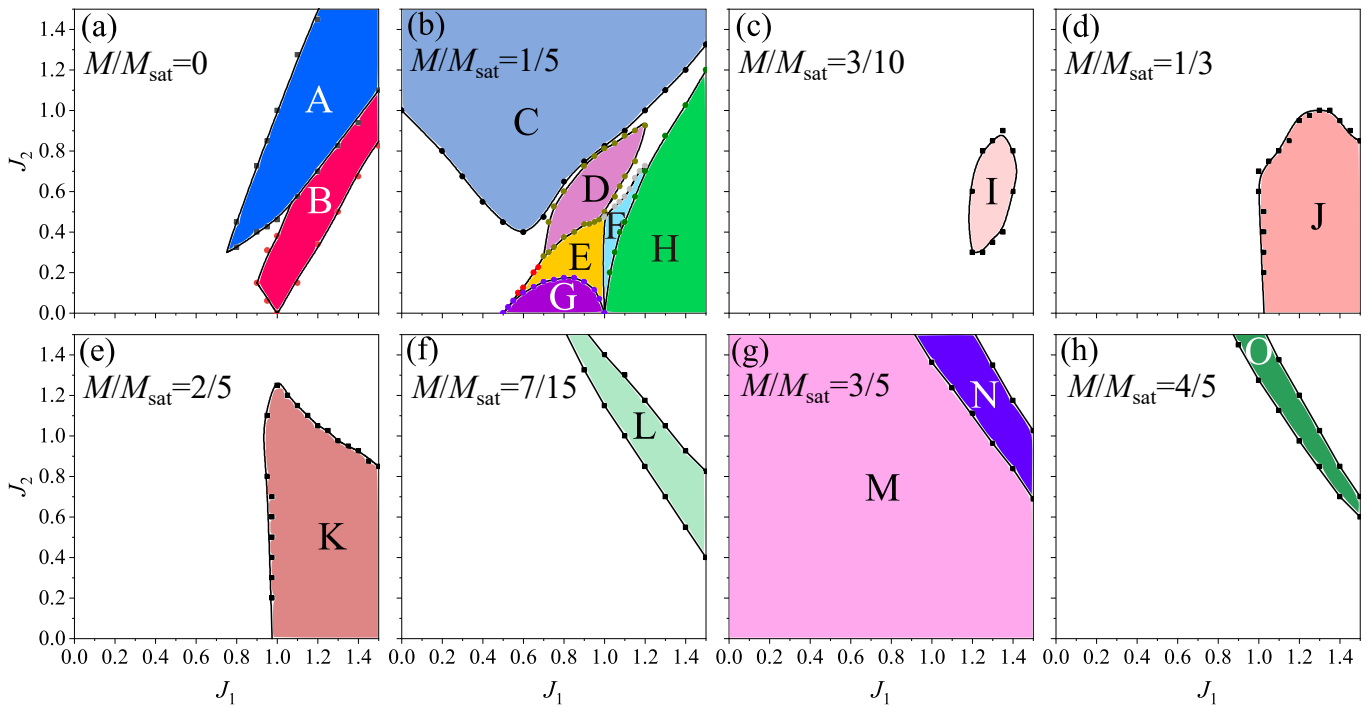}
\caption{Magnetic phase diagram of the KSC at $M/M_{\rm sat}$ = 0 (a), 1/5 (b), 3/10 (c), 1/3 (d), 2/5 (e), 7/15 (f), 3/5 (g), and 4/5 (h).
The colored regions denote the magnetization-plateau phases, which are labeled with letters A to O.
\label{phase}}
\end{figure*} 

In this paper, we construct magnetic phase diagrams under the magnetic field in the spin-1/2 KSC using the density matrix renormalization group (DMRG) method. 
We confirm that magnetization plateaus appear in the wide range of exchange parameters at $M/M_{\rm sat}$ = 0, 3/10, 1/3, 2/5, 7/15, 3/5, and 4/5, in addition to the phase diagram at $M/M_{\rm sat}$ = 1/5 reported elsewhere~\cite{ksc9}.
We find that one of the 3/5 plateau phases has a doubly degenerate entanglement spectrum and edge excitations, similar to the spin-1 Haldane chain. A phase with these characteristics has been identified at  $M/M_{\rm sat}=1/5$~\cite{ksc9}.
Therefore, there is probability that these magnetization plateaus will be observed experimentally in the future using, for example, the model compound~\cite{kscexp}. 

This paper is organized as follows. In Sec.~\ref{sec2}, we describe the KSC model and numerical method. 
In Sec.~\ref{sec3}, we show magnetic phase diagrams of the KSC  and discuss the magnetic properties of the phases separately in subsections. 
Finally, a summary is given in Sec~\ref{sec4}.

\section{model and method}
\label{sec2}

The Hamiltonian for the spin-$\frac{1}{2}$ KSC in a magnetic field is defined as
\begin{eqnarray} 
H &=& \sum_{\langle i,j \rangle }J_{i,j} \mathbf{S}_i \cdot \mathbf{S}_j - h\sum_i S^{z}_i,
\label{Hami}
\end{eqnarray}
where $\mathbf{S}_i$ is the spin-$\frac{1}{2}$ operator at $i$-th site, $S^z_i$ is $z$ component of $\mathbf{S}_i$, 
$\langle i,j \rangle$ runs over the nearest-neighbor spin pairs, $J_{i,j}$ corresponds to one of $J_{\rm X}$, $J_1$, and $J_2$ shown in Fig.~\ref{lattice}, and $h$ is the magnitude of magnetic field applied in the $z$ direction. 
 In the following we set $J_{\rm X}=1$ as the energy unit. 

We calculate the ground states of the KSC up to the number of sites $N=1000$ under the open boundary condition (OBC) using DMRG method.
The number of states $m$ kept in the DMRG calculations are more than 300 and resulting truncation errors are less than $5 \times 10^{-7}$.
We chose the value of $N$ to stabilize each magnetization plateau.
Since the magnetic structure of each magnetization plateau is different, the value of $N$ is set to be different for each phase.

\section{results and discussion}
\label{sec3}
\subsection{Phase diagrams}

We first show the magnetic phase diagrams of the KSC at $M/M_{\rm sat} = 0, 1/5, 3/10, 1/3, 2/5, 7/15, 3/5,$ and 4/5 in Fig.~\ref{phase}.
These are newly obtained phase diagrams except for $M/M_{\rm sat} = 1/5$. 
The phase diagram at $M/M_{\rm sat} = 1/5$ has already been obtained in the previous study~\cite{ksc9}.
For a given $J_1$ and $J_2$, 
we check to see if there are the magnetization plateaus, performing finite-size scaling up to $N=500$ on both edges of the plateau by the DMRG method.
We use the scaling function of the upper (lower) fields in the plateau, $h_{u(l)} = a_{u(l)}(1/N)^2 + b_{u(l)}(1/N) + c_{u(l)}$.
We defined the plateau as the region satisfying $c_u-c_l > 0.005$, where $c_u-c_l$ is the difference of the upper and lower fields at $N\rightarrow\infty$. Therefore, the plateau regions would not be overestimated.
The boundary between phase E and phase D in Fig.~\ref{phase}(b) is determined from the spin structure for $N=1000$ 
and the boundaries between others in Figs.~\ref{phase}(a), \ref{phase}(b), and \ref{phase}(g) are determined at $N=500$.
As a result, we find fifteen magnetization-plateau phases in the KSC.
The colored regions represent the plateau phases, which are labeled with letters A to O. 
The white regions indicate the region where no magnetization plateau appears.
As shown in Fig.~\ref{phase}, the numbers of the magnetization-plateau phases 
at $M/M_{\rm sat} = 0, 1/5, 3/10, 1/3, 2/5, 7/15, 3/5,$ and 4/5 are 
two [Fig.~\ref{phase}(a)], 
six  [\ref{phase}(b)],
one [\ref{phase}(c)],
one [\ref{phase}(d)],
one [\ref{phase}(e)],
one [\ref{phase}(f)],
two [\ref{phase}(g)], and
one [\ref{phase}(h)], respectively.
To the best of our knowledge, it is a first example for a single model to exhibit fifteen different magnetization-plateau phases.
Each phase is characterized by a magnetic structure obtained by the DMRG calculations.
In the following subsections, we discuss the magnetic properties of each phase for a given $M/M_{\rm sat}$ in detail.

\subsection{$M/M_{\rm sat} =0$}
\begin{figure}[tb]
\includegraphics[width=86mm]{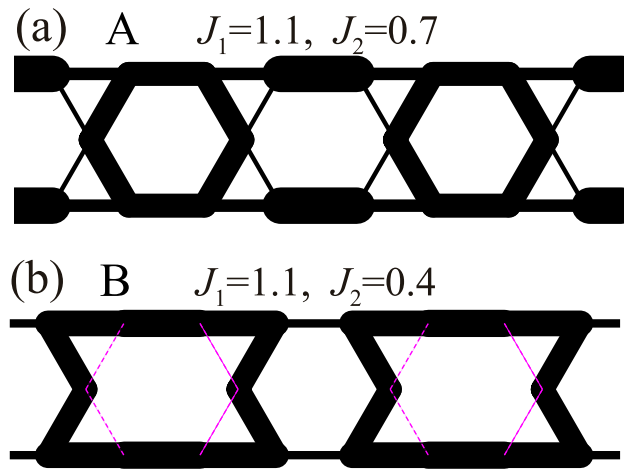}
\caption{The nearest-neighbor spin-spin correlations $\langle \mathbf{S}_i\cdot\mathbf{S}_j\rangle$ around the center of the chain with $N = 200$ under the OBC at $M/M_{\rm sat}=0$
 for phase A at $J_1=1.1$ and $J_2=0.7$ (a) and for phase B at $J_1=1.1$ and $J_2=0.4$ (b). 
Black solid (pink dashed) lines connecting two nearest-neighbor sites denote negative (positive) value of $\langle \mathbf{S}_i\cdot\mathbf{S}_j\rangle$ and their thickness represents the magnitude of the correlation. 
In (a), the value of $\langle \mathbf{S}_i\cdot\mathbf{S}_j\rangle$ at the central horizontal bond is $-0.5774$.
\label{st0}}
\end{figure}

There are two types of the magnetization-plateau phases (A and B) that have been reported~\cite{ksc6}, both of which have a period of $5\times2$ in their magnetic structures shown in Fig.~\ref{st0}.
The black and pink lines and their thickness connecting two nearest-neighbor sites denote the sign and magnitude of the spin-spin correlation $\langle \mathbf{S}_i\cdot\mathbf{S}_j\rangle$, respectively. 
We note that these magnetization plateaus satisfy Oshikawa-Yamanaka-Affleck (OYA) criterion,
which is given by 
\begin{eqnarray} 
pS(1-M/M_{\rm sat})=n,
\label{OYAc}
\end{eqnarray}
where $p$ is a period of magnetic structure in the ground state, $S$ is the spin magnitude, and $n$ is an integer~\cite{OYA}. 
In phases A and B, the OYA criterion is satisfied, because $p=10$~$(=5\times2)$, $S=1/2$, and $M/M_{\rm sat}=0$ give $n=5$.
In phase A, the magnetic structure has dimers and hexamers (hexagon) with strong spin-spin correlations shown in Fig.~\ref{st0}(a),
while in phase B, the magnetic structure has decamers (ten-site-loop) with the strong correlations shown in Fig.~\ref{st0}(b).
These magnetic structures with the strongly coupled clusters (in this case dimer,  hexamer, and decamer) are stabilized by a gain of quantum energies.
As $J_2$ increases, the energy gain of the $J_2$ bond increases, and thus a phase transition from phase B to A occurs as shown in Fig.~\ref{phase}(a).
At $J_1=J_2=1$, phase A has been found in the previous study~\cite{ksc2}. We also obtain phase A at $J_1=J_2=1$. 
Since the spin gap (triplet gap) is very small ($\leq 0.01$) at $J_1=J_2=1$ in the thermodynamic limit,
we consider that a phase boundary between phase A and a gapless phase goes through near $J_1=J_2=1$.

\subsection{$M/M_{\rm sat} =1/5$}
\begin{figure}[tb]
\includegraphics[width=76mm]{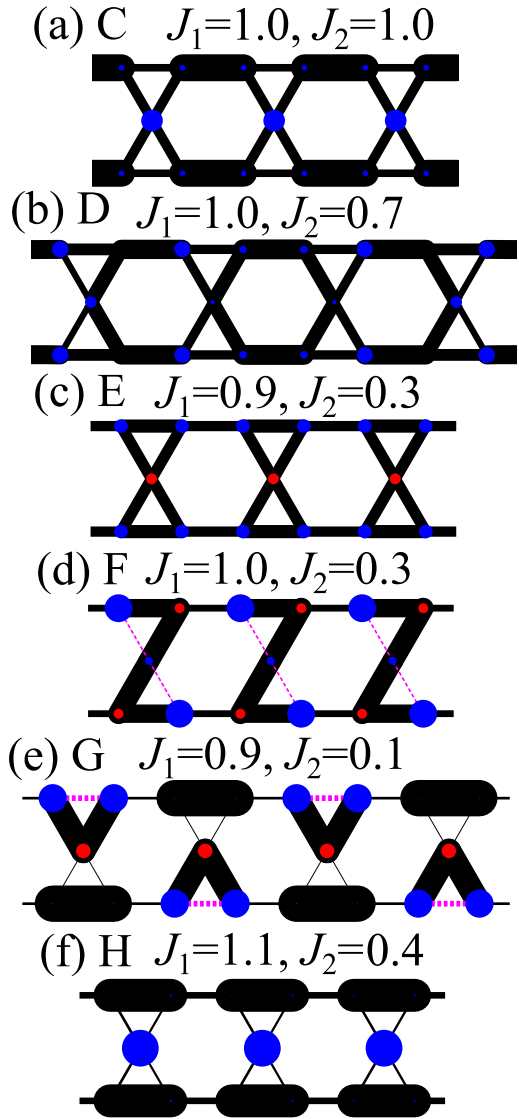}
\caption{The nearest-neighbor spin-spin correlations $\langle \mathbf{S}_i\cdot\mathbf{S}_j\rangle - \langle S_i^z\rangle\langle S_j^z\rangle$ and local magnetizations $\langle S_i^z\rangle$ around the center of the chain with $N = 200$ under the OBC for phases C  (a), D (b), E (c), F (d), G (e), and H (f). 
The values of $J_1$ and $J_2$ shown in each panel refer to the parameters for which DMRG calculations were performed.
Note that phase D is calculated at $M/M_{\rm sat} =1/5+2/N$.
Black solid and pink dashed lines denote the value of $\langle \mathbf{S}_i\cdot\mathbf{S}_j\rangle - \langle S_i^z\rangle\langle S_j^z\rangle$ as in Fig.~\ref{st0}.
Blue (red) circles on each site denote the positive (negative) value of $\langle S_i^z\rangle$, and their diameter represents its magnitude.
In (f), the value of  $\langle S_i^z\rangle$ at the center is 0.4193, and that of $\langle \mathbf{S}_i\cdot\mathbf{S}_j\rangle - \langle S_i^z\rangle\langle S_j^z\rangle$ of the thickest line is $-0.6513$.
\label{st1-5}}
\end{figure} 

There are six types of the magnetization-plateau phases (C to H) that have been reported~\cite{ksc9}.
Figures~\ref{st1-5}(a), \ref{st1-5}(b), \ref{st1-5}(c), \ref{st1-5}(d), \ref{st1-5}(e), and \ref{st1-5}(f) show the nearest-neighbor spin-spin correlation $\langle \mathbf{S}_i\cdot\mathbf{S}_j\rangle - \langle S_i^z\rangle\langle S_j^z\rangle$ 
and local magnetization $\langle S_i^z\rangle$ for phases C, D, E, F, G, and H, respectively.
Note that only phase D is calculated for $M/M_{\rm sat} =1/5+2/N$ due to the presence of edge excitations.
In phases C, E, F, and H, the period of their magnetic structures is $5\times1$ as shown in Figs.~\ref{st1-5}(a), \ref{st1-5}(c), \ref{st1-5}(d), and \ref{st1-5}(f), respectively, while in phase G  the period of its magnetic structure is $5\times2$ as shown in Fig.~\ref{st1-5}(e).
Therefore, the OYA criterion for these five plateaus is satisfied, because $p=5$ for phases C, E, F, and H,  and $p=10$ for phase G give $n=2$ and $n=4$, respectively.

Phases G and H appear in the region where $J_2$ is small.
In the five-site model (i.e., $J_2 = 0$), phase H is obtained in $J_2>1$.
 In other words, the reason for the appearance of phase H can be explained by a zero-order perturbation to $J_2$.
We have named the five-site “$\Xi$”-like structure after its shape of spin correlation as shown in Fig.~\ref{st1-5}(f).
Whereas, the reason for the appearance of phase G in $0.5<J_1<1$ and $J_2 \ll 1$ can be understood by treating $J_2$ as a first-order perturbation~\cite{ksc6}.
Phase C is stabilized by forming singlets by the $J_2$ bonds in $J_2\gg1$, but also appears at $J_2\sim 1$ due to the effect of frustration.
The reasons for the stabilization of phases E and F are currently less understood.

Among the six phases, phase D has the most interesting features.
It has been reported that the magnetic structure is not periodic as shown in Fig.~\ref{st1-5}(b) but has a spin-density-wave-like feature~\cite{ksc9}.
The period of the ground state of phase D is not yet known. Therefore, it is not clear whether phase D satisfies the OYA criterion.
Phase D is a gapless spin liquid 
and its entanglement spectra are doubly degenerate~\cite{ksc9}.
Here, this "gapless" means that there is a gapless excitation in the subspace where the total $S^z$ is fixed.
This means that phase D has the properties of both a half-integer spin chain and a spin-1 Haldane chain.
To describe these properties, a resonating dimer-monomer liquid state has been proposed~\cite{ksc9}.
This phase is distributed over a finite parameter region, not just at $J_1 = 1.0$ and $J_2 = 0.7$ [see Fig.~\ref{phase}(b)].
Therefore, there is probability that phase D will also be discovered experimentally in the future.

\subsection{$M/M_{\rm sat} =3/10$}
\begin{figure}[tb]
\includegraphics[width=86mm]{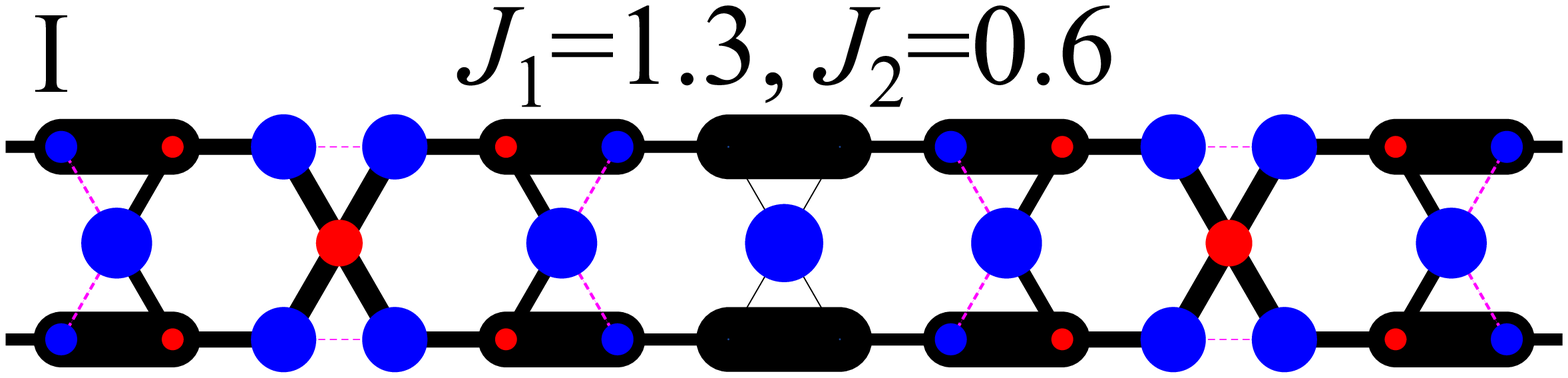}
\caption{The nearest-neighbor spin-spin correlations $\langle \mathbf{S}_i\cdot\mathbf{S}_j\rangle - \langle S_i^z\rangle\langle S_j^z\rangle$ and local magnetizations $\langle S_i^z\rangle$ around the center of the chain with $N = 195$  under the OBC at $M/M_{\rm sat}=3/10+1/2N$ for phase I at $J_1=1.3$ and $J_2=0.6$. 
Black solid and pink dashed lines denote the value of  $\langle \mathbf{S}_i\cdot\mathbf{S}_j\rangle - \langle S_i^z\rangle\langle S_j^z\rangle$, while blue and red circles denote the value of $\langle S_i^z\rangle$, as in Fig.~\ref{st1-5}.
The value of  $\langle S_i^z\rangle$ at the center site is 0.4870, and that of $\langle \mathbf{S}_i\cdot\mathbf{S}_j\rangle - \langle S_i^z\rangle\langle S_j^z\rangle$ for the $J_1$ bond at the center is $-0.6989$.
\label{st3-10}}
\end{figure}

The magnetic structure of this plateau phase (phase I) has a period of four-unit cells ($5\times4$ sites) as shown in Fig.~\ref{st3-10}. 
This plateau satisfies the OYA criterion because 
$p=20$, $S=1/2$, and $M/M_{\rm sat}=3/10$ give $n=7$ in Eq.~(\ref{OYAc}).
The width of the magnetization plateau has been found to be very narrow in the previous study~\cite{ksc6}.  
In the present study, however, it is found that the region showing the 3/10 plateau is wider as shown in Fig.~\ref{phase}(c). 
In this region, $J_1$ is larger than $J_2$ and $J_{\rm X} (=1)$, and $J_2$ is less than $J_{\rm X}$.
As shown in Fig.~\ref{st3-10}, the magnetic structure has a shape like $\cdots\Xi$X$\Xi\Xi\Xi$X$\Xi\cdots$, 
where the five-site “X”-like structure has $M\sim3/2$ and the five-site “$\Xi$”-like structure has $M\sim1/2$.
We note that when investigating phase I under the OBC, the total number of site needs to be fixed to $N=5\times(4n_p+3)$, where $n_p$ is a natural number, because the left end of the chain exhibits a $\Xi$X$\Xi\Xi\Xi\cdots$ structure and 
the right end exhibits a $\cdots\Xi\Xi\Xi$X$\Xi$ structure.
Therefore, the plateau of phase I appears at $M/M_{\rm sat}=3/10+1/2N$.
In the thermodynamic limit $N\rightarrow\infty$, this plateau appears at $M/M_{\rm sat} =3/10$.
The spin-spin correlations of the $J_1$ bond in the five-site ``$\Xi$''-like structures are negatively large. 
In addition, those of $J_{\rm X}$ bonds in the five-site ``X''-like structures are reasonably large.
Since the spin-spin correlations between the ``X''-like structure and the ``$\Xi$''-like structure and between the ``$\Xi$''-like structures are small as shown in Fig.~\ref{st3-10}, the effect of the $J_2$ bonds on the total energy is small.
These explain the presence of phase I in the region $J_1>J_{\rm X} >J_2$.

\subsection{$M/M_{\rm sat} =1/3$} 

\begin{figure}[tb]
\includegraphics[width=86mm]{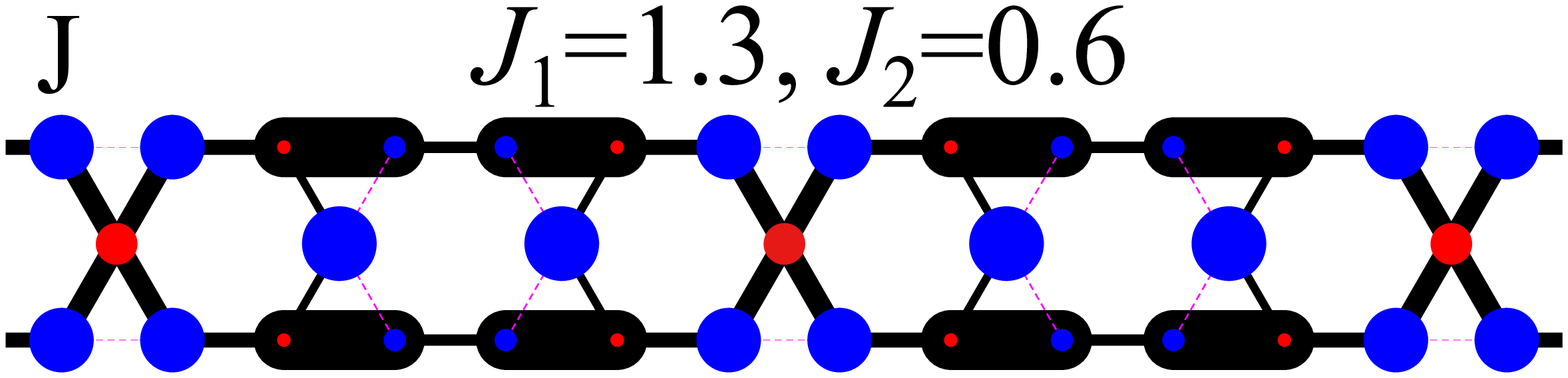}
\caption{The nearest-neighbor spin-spin correlations $\langle \mathbf{S}_i\cdot\mathbf{S}_j\rangle - \langle S_i^z\rangle\langle S_j^z\rangle$ and local magnetizations $\langle S_i^z\rangle$ around the center of the chain with $N = 195$ under the OBC at $M/M_{\rm sat}=1/3$ for phase J at $J_1=1.3$ and $J_2=0.6$. 
Black solid and pink dashed lines denote the value of  $\langle \mathbf{S}_i\cdot\mathbf{S}_j\rangle - \langle S_i^z\rangle\langle S_j^z\rangle$, while blue and red circles denote the value of $\langle S_i^z\rangle$,  as in Fig.~\ref{st1-5}.
The value of  $\langle S_i^z\rangle$ at the center site is $-0.2594$, and that of $\langle \mathbf{S}_i\cdot\mathbf{S}_j\rangle - \langle S_i^z\rangle\langle S_j^z\rangle$ for the $J_{\rm X}$ bond at the center is $-0.2340$.
\label{st1-3}}
\end{figure} 

The magnetic structure of this plateau phase (phase J) has a shape like $\cdots$X$\Xi\Xi$X$\Xi\Xi$X$\cdots$ which has a period of three-unit cells ($5\times3$ sites) as shown in Fig.~\ref{st1-3}. 
In this plateau, since  $p=15$ gives $n=5$ in Eq.~(\ref{OYAc}),  the OYA criterion is satisfied.
The magnetization-plateau width has been found to be narrow in the previous study~\cite{ksc6}, as in the case of phase I.
The present results show that the parameter region with respect to $J_1$ and $J_2$ showing the magnetization plateau is reasonably large as show in Fig.~\ref{phase}(d).
As in the case of phase I, the region satisfies the relationship $J_1>J_{\rm X}>J_2$.
Similar to phase I, this is explained by the fact that the spin-spin correlations of the $J_1$ bonds in the five-site ``$\Xi$''-like  structures are the largest followed by the next largest $J_{\rm X}$ bonds in the five-site ``X''-like structures and the $J_2$ bonds with smaller correlation as shown in Fig.~\ref{st1-3}.

\subsection{$M/M_{\rm sat} =2/5$}
\begin{figure}[tb]
\includegraphics[width=86mm]{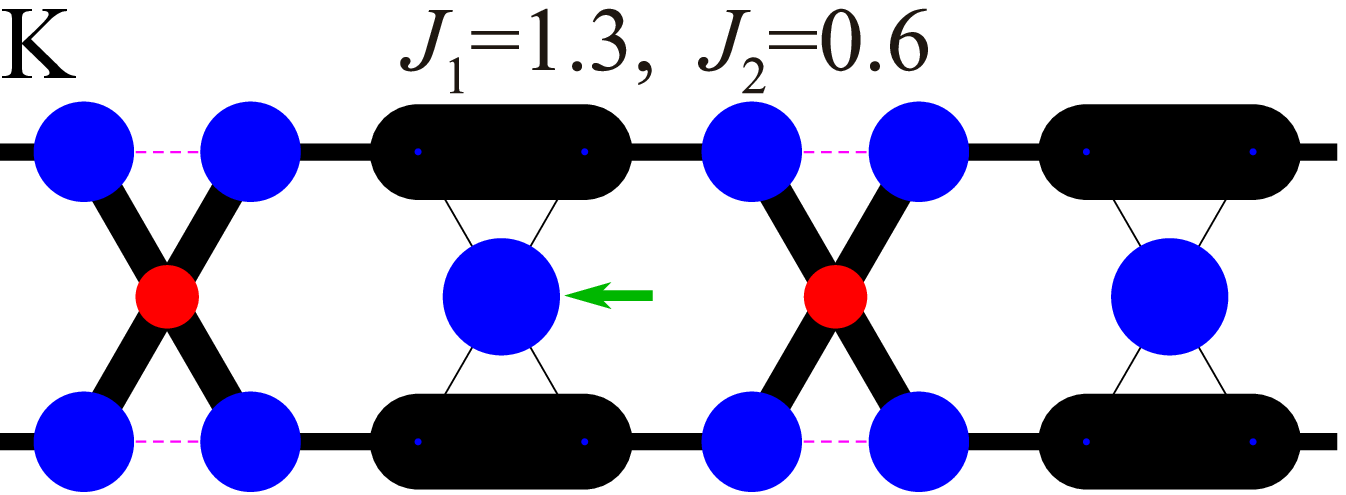}
\caption{The nearest-neighbor spin-spin correlations $\langle \mathbf{S}_i\cdot\mathbf{S}_j\rangle - \langle S_i^z\rangle\langle S_j^z\rangle$ and local magnetizations $\langle S_i^z\rangle$ around the center of the chain with $N = 195$ under the OBC at $M/M_{\rm sat}=2/5+1/N$ for phase K at $J_1=1.3$ and $J_2=0.6$. 
Black solid and pink dashed lines denote the value of  $\langle \mathbf{S}_i\cdot\mathbf{S}_j\rangle - \langle S_i^z\rangle\langle S_j^z\rangle$, while blue and red circles denote the value of $\langle S_i^z\rangle$ as in Fig.~\ref{st1-5}.
The value of  $\langle S_i^z\rangle$ at the site marked with a green arrow is 0.4864, 
and the value of $\langle \mathbf{S}_i\cdot\mathbf{S}_j\rangle - \langle S_i^z\rangle\langle S_j^z\rangle$ of the thickest line is $-0.6871$.
\label{st2-5}}
\end{figure}

The magnetic structure of this plateau phase (phase K) has an alternating order of the five-site ``X''-like and ``$\Xi$''-like structures as shown in Fig.~\ref{st2-5}, which means that the magnetic structure has a period of two-unit cells ($5\times2$ sites).
This plateau satisfies the OYA criterion because of $p=10$ and $n=3$ in Eq.~(\ref{OYAc}).
In $J_2 \ll 1$, it has been found that this model is equivalent to the antiferromagnetic Ising model by the analysis of the first-order perturbation theory~\cite{ksc6}. 
The magnetic structure of a shape $\cdots$X$\Xi$X$\Xi$$\cdots$ in the KSC corresponds to $\cdots\uparrow \downarrow \uparrow \downarrow \cdots$ in the Ising model. 
We note that when investigating phase K under the OBC, the total number of site needs to be fixed to $N=5\times(2n_p+1)$, where $n_p$ is a natural number, because the left end of the chain exhibits a X$\Xi$X$\Xi\cdots$ structure and 
the right end exhibits a $\cdots\Xi$X$\Xi$X structure.
Therefore, the plateau of phase K appears at $M/M_{\rm sat}=2/5+1/N$.
In the thermodynamic limit $N\rightarrow\infty$, this plateau appears at $M/M_{\rm sat} =2/5$.
The stability of phase K has been explained by the first-order perturbation theory for $J_2$ ($\ll J_{\rm X}$)~\cite{ksc6}. 
However, as shown in Fig.~\ref{phase}(e), we find that phase K appears even at $J_2 \sim J_1 \sim 1$, which is beyond the first-order perturbation theory.

Here, we discuss phases I, J, and K.
These three phases have in common that they are composed of the five-site ``X''-like and ``$\Xi$''-like structures.
The magnetic structures of phases I, J, and K have a period of four-unit cells, three-unit cells, and two-unit cells, respectively.
The region showing the plateau becomes larger in the order of phases I, J, and K, as shown in Figs~\ref{phase}(c), \ref{phase}(d), and \ref{phase}(e).
In addition, no plateau phase with a periodic structure with more than four-unit cells has been found in KSC.
These are because the longer the period is, the more unstable the structure becomes.

\subsection{$M/M_{\rm sat} =7/15$}
\begin{figure}[tb]
\includegraphics[width=86mm]{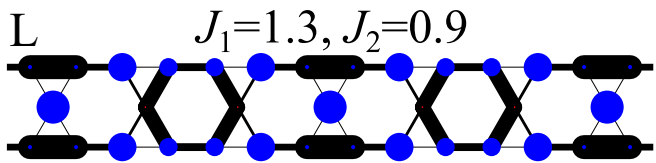}
\caption{The nearest-neighbor spin-spin correlations $\langle \mathbf{S}_i\cdot\mathbf{S}_j\rangle - \langle S_i^z\rangle\langle S_j^z\rangle$ and local magnetizations $\langle S_i^z\rangle$ around the center of the chain with $N = 235$ under the OBC at $M/M_{\rm sat} =7/15+4/3N$ for phase L at $J_1=1.3$ and $J_2=0.9$. 
Black solid lines denote the value of  $\langle \mathbf{S}_i\cdot\mathbf{S}_j\rangle - \langle S_i^z\rangle\langle S_j^z\rangle$, 
while blue and red circles denote the value of $\langle S_i^z\rangle$ as in Fig.~\ref{st1-5}.
The value of  $\langle S_i^z\rangle$ at the center site is $0.4976$, and that of $\langle \mathbf{S}_i\cdot\mathbf{S}_j\rangle - \langle S_i^z\rangle\langle S_j^z\rangle$ for $J_1$ bond at the center is $-0.6253$.
\label{st7-15}}
\end{figure}

The magnetic structure of this plateau phase (phase L) consists of almost isolated monomers with $M\sim1/2$, dimers with $M\sim0$, and hexamers with $M\sim1$ as reported preciously~\cite{ksc6}.
The period of this magnetic structure is $5\times3$ as shown in Fig.~\ref{st7-15}.
This plateau also satisfies the OYA criterion because $p=15$ gives $n=4$ in Eq.~(\ref{OYAc}).
We note that when investigating phase L under the OBC, the total number of site needs to be fixed to $N=5\times(3n_p+2)$, where $n_p$ is a natural number, because the ten sites (two units) at both ends of the chain exhibit the four monomers and hexagon.
Therefore, the plateau of phase L appears at $M/M_{\rm sat}=7/15+4/3N$.
In the thermodynamic limit $N\rightarrow\infty$, this plateau appears at $M/M_{\rm sat} =7/15$.
The region showing the plateau  does not have the condition $J_1>1$ and $J_2\lesssim1$, unlike phases I, J, and K.
Figure~\ref{phase}(f) shows that the region is distributed along the line $J_1\sim-1.4J_2+2.75$.
This is probably because the $J_2$ and $J_{\rm X}$ bonds have the energy gain in the hexamers with the large correlations, and the $J_1$ bonds also have the energy gain in the dimers with the large correlation.
Because of the energy gain in both $J_1$ and $J_2$ bonds, the region showing the plateau of phase L is not only  $J_2 > J_1$  but also $J_2 < J_1$.

\subsection{$M/M_{\rm sat} =3/5$}
\begin{figure}[tb]
\includegraphics[width=86mm]{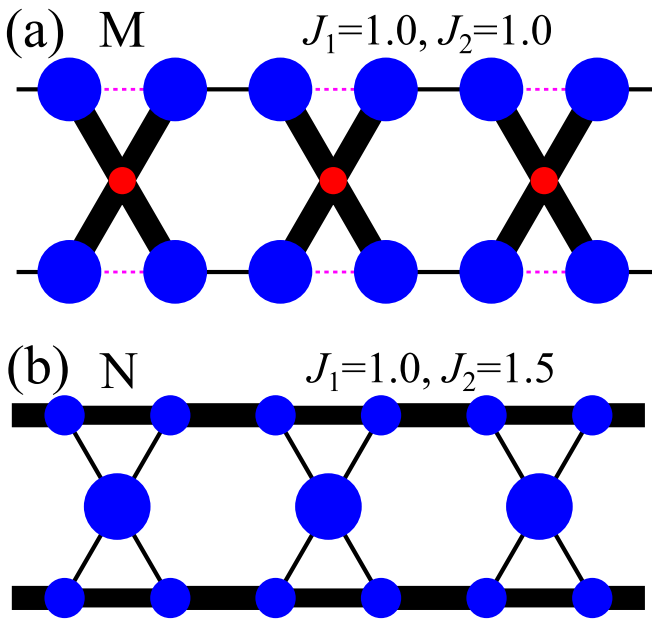}
\caption{The nearest-neighbor spin-spin correlations $\langle \mathbf{S}_i\cdot\mathbf{S}_j\rangle - \langle S_i^z\rangle\langle S_j^z\rangle$ and local magnetizations $\langle S_i^z\rangle$ around the center of the chain with $N = 200$ under the OBC for phase M at $J_1=1.0$ and $J_2=1.0$ at $M/M_{\rm sat} =3/5$ (a) and for phase N at $J_1=1.0$ and $J_2=1.5$ at $M/M_{\rm sat} =3/5+2/N$ (b).
Black solid and pink dashed lines denote the value of  $\langle \mathbf{S}_i\cdot\mathbf{S}_j\rangle - \langle S_i^z\rangle\langle S_j^z\rangle$, 
while and blue and red circles denote the value of $\langle S_i^z\rangle$, respectively, as in Fig.~\ref{st1-5}.
In (a), the value of  $\langle S_i^z\rangle$ at the center site is $-0.1791$, and that of $\langle \mathbf{S}_i\cdot\mathbf{S}_j\rangle - \langle S_i^z\rangle\langle S_j^z\rangle$ for $J_{\rm X}$ bond at the center is $-0.2727$.
\label{st3-5}}
\end{figure}

There are two types of magnetization plateau phases (M and N), both of which have a period of $5\times1$ in their magnetic structures shown in Fig.~\ref{st3-5}.
In these plateaus, the OYA criterion is satisfied, because $p=5$ gives $n=1$ in Eq.~(\ref{OYAc}).
Within the scope of our investigation, the region showing phase N is distributed along the line $J_1\sim-0.6J_2+2$.
The appearance of phase M in $J_2, J_1 \ll 1$ is easily understood by analyzing the five sites (single unit)~\cite{ksc6}.
Moreover,  in the classical spin system, this magnetization plateau corresponds to an up-up-down-up-up structure.
Therefore, phase M is very stable and thus the region showing phase M is very large as shown in Fig.~\ref{phase}(g).
Phase M has already been found in our previous study~\cite{ksc6}.
By contrast, phase N is firstly found in the present study.
Phase N is investigated in detail in the following.

\begin{figure}[tb]
\includegraphics[width=86mm]{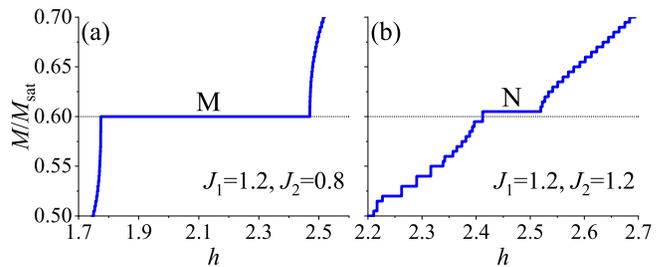}
\caption{Magnetization curves around $M/M_{\rm sat}=3/5$ of the KSC with $N=400$ under the OBC for phase M at $J_1=1.2$ and $J_2=0.8$ (a), and for phase N at $J_1=1.2$ and $J_2=1.2$ (b).
The horizontal dotted lines represent $M/M_{\rm sat}=3/5$.
\label{M-H}}
\end{figure}

Figures~\ref{M-H}(a) and~\ref{M-H}(b) show magnetization curves with $N=400$ under the OBC at $J_1=1.2$ and $J_2=0.8$ (phase M), and $J_1=1.2$ and $J_2=1.2$ (phase N), respectively. 
The magnetization plateau of phase M appears at $M/M_{\rm sat} =3/5$, whereas that of phase N appears at $M/M_{\rm sat} =3/5+2/N$ that is only one step larger than $3/5$.
This $2/N$ implies edge excitations that appear in the spin-1 Haldane chain under the OBC~\cite{ER}.
In the following, we regard the magnetization of phase N under the OBC as $M/M_{\rm sat}=3/5+2/N$.

\begin{figure}[tb]
\includegraphics[width=86mm]{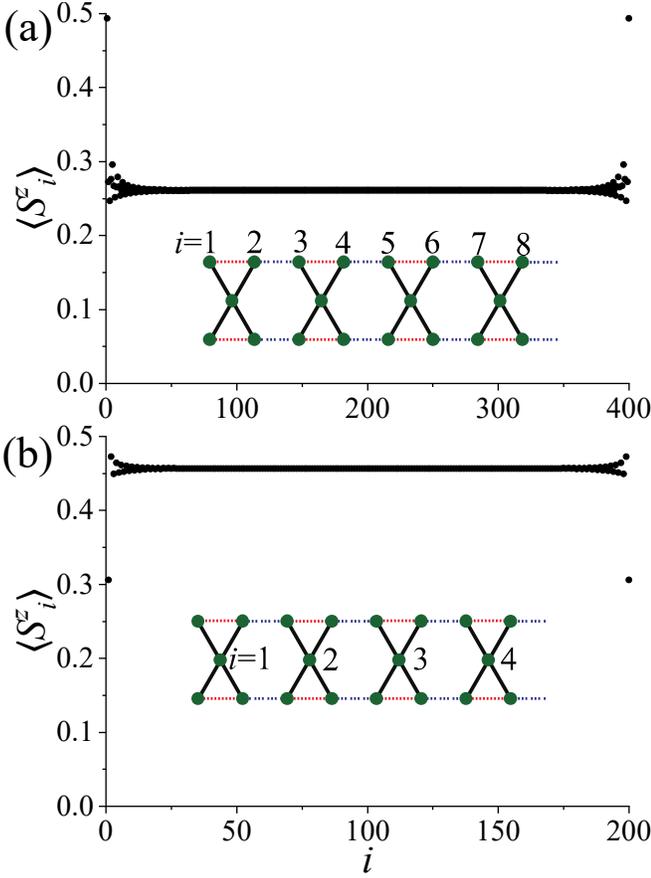}
\caption{ Local magnetization $\langle S^z_i\rangle$ at sites along the upper edge (a) and at the central sites (b) of the KSC with $M/M_{\rm sat} = 3/5+2/N$ under the OBC for $N=1000$ at $J_1=1.2$ and $J_2=1.3$.
Here, $i$ represents the position of each site, labeled in the inset.
\label{locsz}}
\end{figure}

In order to confirm  edge excitations in phase N, we investigate local magnetization.
Figures~\ref{locsz}(a)~and~\ref{locsz}(b) show the local magnetization $\langle S_i^z \rangle$ at sites along the upper edge and at central sites, respectively, in the KSC with $N=1000$ under the OBC.
Both ends of the upper edge ($i = 1$ and $i = 400$) have almost a full moment $\langle S^z_{i=1(400)}\rangle=0.4932$,
 corresponding to an edge excitation, which leads to the term $2/N$ in $M/M_{\rm sat} =3/5+2/N$.
Therefore, we conclude that there are edge excitations in phase N under the OBC.

\begin{figure}[tb]
\includegraphics[width=86mm]{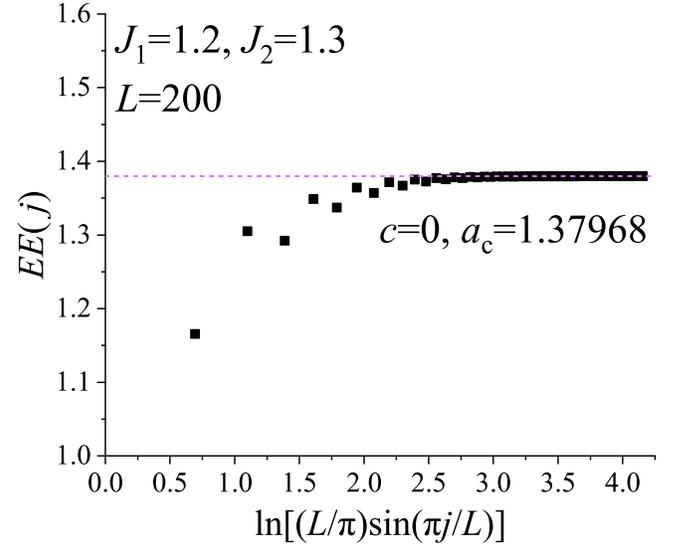}
\caption{Entanglement entropy of the KSC with $L=200$ ($N=5\times200$) at $M/M_{\rm sat}=3/5+2/N$ under the OBC at $J_1=1.2$ and $J_2=1.3$, where $L$ is the number of five-site units in the KSC and $j$ denotes the position of the five-site unit. 
The entanglement entropy $EE(j)$ is plotted as a function of $\ln\left[(L/\pi) \sin \left( \pi j/L \right) \right]$.
The pink dashed line represents the line with $c=0$ and $a_{\rm c}$=1.37968 in Eq.~(\ref{EEeq}).
\label{EE}}
\end{figure}

Next, to confirm that phase N is equivalent to the Haldane phase in the spin-1 Haldane chain, 
we calculate the entanglement entropy.
The relationship between the position-dependent entanglement entropy, $EE(j)$, and the central charge $c$ characterizing excitation properties~\cite{EE1} satisfies 
\begin{eqnarray} 
EE(j) &=& \frac{c}{b_{\rm c}}\ln\left[\frac{L}{\pi} \sin \left( \frac{\pi j}{L} \right) \right] + a_{\rm c},
\label{EEeq}
\end{eqnarray}
where $a_{\rm c}$ is a nonuniversal constant, $b_{\rm c} = 6 $ (3) for the OBC (PBC)~\cite{EE2,EE3,EE4,EE5}, $L$ is the number of five-site units in the KSC, and $j$ denotes the position of the five-site unit.
When the spin-spin correlation decays exponentially and an energy gap (spin gap) is finite, e.g., in the Haldane phase, the value of $c$ is zero.
Figure~\ref{EE} shows the entanglement entropy $EE(j)$ with respect to $\ln\left[(L/\pi) \sin \left( \pi j/L \right) \right]$.
The pink dashed line represents the straight line with $c=0$ and $a_{\rm c}$=1.37968 in Eq.~(\ref{EEeq}).
This straight line and solid squares calculated by DMRG in $\ln\left[(L/\pi) \sin \left( \pi j/L \right) \right] > 2.0$ almost agree.
Therefore, $c$ in phase N is zero.
We conclude that an energy gap exists in phase N as in the case of the Haldane phase.

\begin{figure}[tb]
\includegraphics[width=86mm]{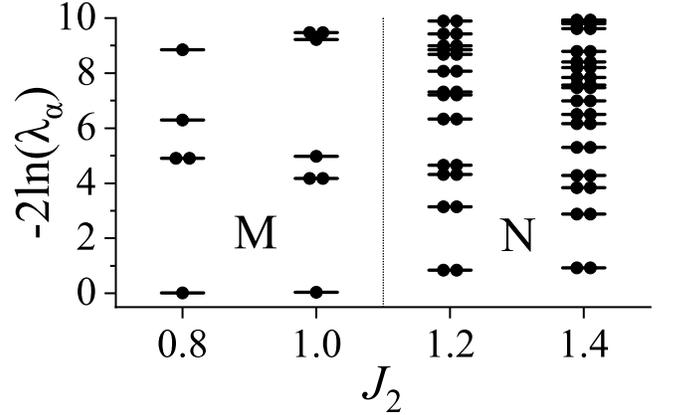}
\caption{Entanglement spectra of the KSC for $L=80 (N=5\times80)$ with respect to $J_2$ at $J_1$ =1.2,
where $L$ is the number of five-site units in the KSC. 
The horizontal bars and filled circles indicate the values of  $\lambda_\alpha$ and degree of degeneracy of $\lambda_\alpha$, respectively.
M and N in the figure represent the phases.
The vertical dotted line represents the phase boundary between M and N.
\label{ES}}
\end{figure}

Next, we calculate the entanglement spectrum of phases M and N.
The ground state, $\left|0\right>$, can be Schmidt decomposed as follows:
\begin{equation} 
\left|0\right>=\sum_{\alpha}\lambda_\alpha\left|\Phi^L_\alpha\right>\left|\Phi^R_\alpha\right>,
\end{equation}
where $\left|\Phi^L_\alpha\right>$ and $\left|\Phi^R_\alpha\right>$ are orthonormal basis
vectors of the left and right part of the system, respectively~\cite{QE,ES1}.
Here, $\lambda_\alpha^2$ are the eigenvalues of the reduced density matrix and $\lambda_\alpha>0$.
The entanglement spectra are defined as $-2\ln(\lambda_\alpha)$.
In the present study, we obtain the reduced density matrix using the DMRG calculation.
Figure~\ref{ES} shows the results of the entanglement spectrum for phases M and N.
In phase M, the entanglement spectrum with the smallest value is non-degenerate.
By contrast, in phase N, all entanglement spectra are doubly degenerate.
These results do not change with the value of $J_2$ as shown in Fig.~\ref{ES}.
The doubly degenerate entanglement spectra under the OBC are identical to the feature of the spin-1 Haldane chain~\cite{ES1}. 

To conclude that phase N is equivalent to the Haldane phase, we further investigate a parameter set of $J_1 =1.95$ and $J_2=0.1$,
where the $J_2$ term can be treated as a perturbation due to $J_1>J_{\rm X}\gg J_2$.
We confirm that even at $J_1 =1.95$ and $J_2=0.1$, a magnetization plateau appears at $M/M_{\rm sat} =3/5+2/N$, $c$ is zero, and the entanglement spectra are doubly degenerate (the results are not shown here).
From these results, 
phase N is found to be continuous from the region shown in Fig.~\ref{phase}(g) to $J_1\sim2$ and $J_2\sim0$.
The perturbation theory for the KSC at $J_1\sim2$ and $J_2 \ll 1$ has already been reported~\cite{PT}.
When the first-order perturbation calculation is performed, the KSC at $J_1\sim2$ and $J_2 \ll 1$ is equivalent to the spin-1 Haldane chain.
Therefore, we conclude that phase N is equivalent to the spin-1 Haldane phase. 
We note that a similar Haldane phase in a magnetic field has been found in the 1/3 plateau of the ferromagnetic-ferromagnetic-antiferromagnetic chain~\cite{HH1,HH2}.
However, our KSC differs in that there is no ferromagnetic interaction and the unit cell contains five sites, not three sites.

\subsection{$M/M_{\rm sat} =4/5$}

\begin{figure}[tb]
\includegraphics[width=86mm]{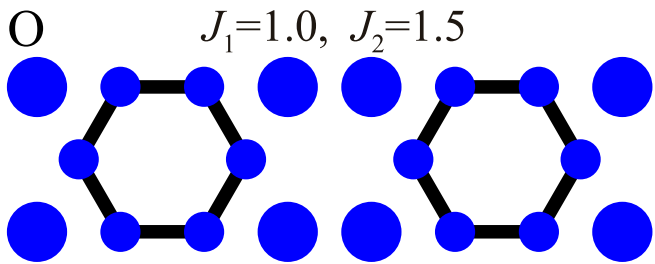}
\caption{
The nearest-neighbor spin-spin correlations $\langle \mathbf{S}_i\cdot\mathbf{S}_j\rangle - \langle S_i^z\rangle\langle S_j^z\rangle$ and local magnetizations $\langle S_i^z\rangle$ around the center of the chain with $N = 200$ under the OBC at $M/M_{\rm sat} = 4/5$ for phase O at $J_1=1.0$ and $J_2=1.5$. 
Black solid lines and blue circles denote the value of  $\langle \mathbf{S}_i\cdot\mathbf{S}_j\rangle - \langle S_i^z\rangle\langle S_j^z\rangle$ and $\langle S_i^z\rangle$, respectively, as in Fig.~\ref{st1-5}.
The value of  $\langle S_i^z\rangle$ of the largest circle is $0.5$, and that of $\langle \mathbf{S}_i\cdot\mathbf{S}_j\rangle - \langle S_i^z\rangle\langle S_j^z\rangle$ for a hexagon edge is $-0.1944$.
\label{st4-5}}
\end{figure} 

The magnetic structure of this plateau phase (phase O) has a period of two-unit cells ($5\times2$ sites) as shown in Fig.~\ref{st4-5}.
This plateau also satisfies the OYA criterion because $p=10$ gives $n=1$ in Eq.~(\ref{OYAc}).
The presence of an exact state of this magnetization plateau at $J_1=1$ and $J_2=1.5$ has been revealed in the previous studies~\cite{ksc3,ksc6}.
This is also understood from the fact that the nearest-neighbor spin-spin correlations $\langle \mathbf{S}_i\cdot\mathbf{S}_j\rangle - \langle S_i^z\rangle\langle S_j^z\rangle$ between hexagons and up spins with full moment is zero, as shown in Fig.~\ref{st4-5}.
This exact state has been also found in $M/M_{\rm sat}=7/9$ plateau of the two-dimensional kagome lattice. 
Figure~\ref{phase}(h) shows that the region of phase O is distributed along the line $J_1\sim -1.6J_2 + 3.0$.
Therefore, phase O exists in a finite region, not only at $J_1=1$ and $J_2 = 1.5$.

\section{summary}
\label{sec4}
Inspired by recent researches for the KSCs with a lattice structure similar to the kagome lattice, 
we investigated the ground state of the spin-1/2 Heisenberg model on the KSC in a magnetic field using the DMRG method.
We found fifteen types of magnetization-plateau phases.
This means that the KSC has strong frustration effects similar to the kagome lattice.
In addition, we constructed the phase diagrams at $M/M_{\rm sat} = 0, 1/5, 3/10, 1/3, 2/5, 7/15, 3/5$, and 4/5.
All phases we found have finite regions with respect to $J_1$ and $J_2$.
The most important result of the present study is found that one of the phases at $M/M_{\rm sat} = 3/5$ is equivalent to the spin-1 Haldane phase.
We consider that these plateaus, except for O phase, are not related to the theoretically predicted plateaus at $M/M_{sat} = 0, 1/9, 1/3,$ and 5/9 in the two-dimensional kagome lattice.
However, we believe that KSC has more phases than kagome lattice due to strong quantum fluctuations inherent in one dimension.
The KSC compounds with five exchange interactions have been previously reported~\cite{kscexp}.
Therefore, we expect that at least one of the fifteen magnetization plateaus may be observed in experiments in the future. 
We hope that the KSC will be studied experimentally and the plateau phases will be observed. 
A future theoretical study would be extended to investigate the model with five exchange interactions or ferromagnetic interactions (e.g. $J_{\rm X}=-1$).

\begin{acknowledgments}
This work was supported in part by MEXT as a social and scientific priority issue (Creation of new functional devices and high-performance materials to support next-generation industries (CDMSI) to be tackled by using post-K computer and by MEXT HPCI Strategic Programs for Innovative Research (SPIRE) (hp190198). The numerical calculation was partly carried out at the K Computer, Institute for Solid State Physics, The University of Tokyo, and the Information Technology Center, The University of Tokyo. 
This work is also supported by Grants-in-Aid for Scientific Research (19H01829, JP19H05825, 17K14148 and 21H03455) from MEXT, Japan , and by JST PRESTO (Grant No. JPMJPR2013).
\end{acknowledgments}

\end{document}